\documentclass[journal]{IEEEtran}
\usepackage{cite}
\usepackage{graphicx}
\usepackage{amsmath}
\usepackage{multirow}
\usepackage{url}
\usepackage{xcolor}
\usepackage{color}
\usepackage{soul}
\begin{document}
\title{A Computational Efficient Pumped Storage Hydro Optimization in the Look-ahead Unit Commitment and Real-time Market Dispatch Under Uncertainty}

\author{Bing Huang,~\IEEEmembership{Member,~IEEE,}
	  Arezou Ghesmati,~\IEEEmembership{Member,~IEEE,}
	  Yonghong~Chen,~\IEEEmembership{Fellow,~IEEE,}
        and~Ross~Baldick,~\IEEEmembership{Fellow,~IEEE,}
        % <-this % stops a space
\thanks{Bing Huang, Arezou Ghesmati and Y. Chen are with Midcontinent Independent System Operator, Inc. (MISO), Carmel, IN, 46032 USA.}% <-this % stops a space
\thanks{R. Baldick is an emeritus professor at the Department
of Electrical and Computer Engineering, University of Texas, Austin, TX, 78712 USA.}
\thanks{This work has been submitted to the IEEE for possible publication. Copyright may be transferred without notice, after which this version may no longer be accessible.}
}


\markboth{Journal of \LaTeX\ Class Files,~Vol.~14, No.~8, August~2015}%
{Shell \MakeLowercase{\textit{et al.}}: Bare Demo of IEEEtran.cls for IEEE Journals}



\maketitle

\begin{abstract}
Pumped storage hydro units (PSHU) are great sources of flexibility in power systems. This is especially valuable in modern systems with increasing shares of intermittent renewable resources. However, the flexibility from PSHUs, particularly in the real-time market, has not been thoroughly studied. The storage optimization in a real-time market hasn't been well addressed. To enhance the use of PSH resources and leverage their flexibility, it is important to incorporate the uncertainties, properly address the risks and avoid increasing too much computational burdens in the real-time market operation. To provide a practical solution to the daily operation of a PSHU in a single day look-ahead commitment (LAC) and real-time market, this paper proposes two pumped storage hydro (PSH) models that only use probabilistic price forecast to incorporate uncertainties and manage risks in the LAC and real-time market operation. The price forecast scenarios are formulated only on PSHUs that minimizes the computational challenges to the Security Constrained Unit Commitment (SCUC) problem. Numerical studies in Mid-continent Independent System Operator (MISO) demonstrate that the proposed models improves market efficiency. Compared to traditional stochastic and robust unit commitment, the proposed methods only moderately increase the solving time from current practice of deterministic LAC. Probabilistic forecast for Real Time Locational Marginal Price (RT-LMP) on PSH locations is created and embedded into the proposed stochastic optimization model, an statistical robust approach is used to generate scenarios for reflecting the temporal inter-dependence of the LMP forecast uncertainties.

%A stochastic PSH model is developed using the probabilistic price forecasts to capture the real-time market uncertainties. In addition, a robust PSH model is developed to manage the financial risk for a PSHU in the real-time market.

\end{abstract}


% \fi
\IEEEpeerreviewmaketitle

\section*{Nomenclature}

Indices and sets:

%\makebox[1.5cm][l]{$i \in I_{VM}$} buses with virtual biddings;

%\makebox[1.5cm][l]{$i \in I_{D}$} buses with dispatchable demands;
\begin{IEEEdescription}[\IEEEusemathlabelsep\IEEEsetlabelwidth{$V_1,V_2.5$}]
	\setlength{\itemsep}{2pt}
	\item[$t \in \mathcal{T}$] time $t$ in the set of time intervals;
	\item[$g \in \mathcal{G}_{psh}$] unit $g$ in the set of all PSHUs in the system;
	\item[$g \in \mathcal{G}_{psh,r}$] unit $g$ in the set of PSHUs that share the same reservoir $r$;
	\item[$g \in \mathcal{G}$] unit $g$ in the set of the generating units besides PSHUs in a system;
	\item[$s \in \mathcal{S}$] scenario $s$ in the set of probabilistic scenarios;
	\item[$r \in \mathcal{R}$] reservoir $r$ in the set of reservoirs.
\end{IEEEdescription}

Data [units]:

%\makebox[1.5cm]{$C_g^{NL,m}$} no load cost of configuration $m$ of unit $g$ [\$/hr];

%\makebox[1.5cm]{$C_g^{TR,mn}$} transition cost between configuration $m$ and $n$ of unit $g$ [\$/hr];

%\makebox[1.5cm]{$C_{gs}^{TR,0n}$} transition cost of type $s$ between configuration $0$ and $n$ of unit $g$ [\$/hr];

\begin{IEEEdescription}[\IEEEusemathlabelsep\IEEEsetlabelwidth{$V_1,V_2.5$}]
	\setlength{\itemsep}{2pt}
	\item[$D_t$] system net load at period $t$ [MW];
	%\textcolor{blue}{\item[$D^{DA}_t$] deterministic system net load forecast at period $t$ made for day-ahead market [MW];}
	\item[$\eta_g^{gen}$] generating efficiency of the PSHU $g$ [NA];
	\item[$\eta_g^{pump}$] pumping efficiency of the PSHU $g$ [NA];
	\item[$E_{r,t_1}$] initial energy level of the reservoir $r$ [MWh];
	\item[$E_{r,T+1}$] final energy level of the reservoir $r$ [MWh];
	\item[$\overline{E_r}$] maximum energy level of the reservoir $r$ [MWh];
	\item[$\underline{E_r}$] minimum energy level of the reservoir $r$ [MWh];
	\item[$P_s$] weighted probability of scenario $s$ [NA];
	\item[$LMP_{g,s,t}^{t_0}$] Location Marginal Price forecast made at $t_0$ for unit $g$ in scenario $s$ at time $t$ [\$/MW];
	\item[$Q^{gen,DA}_{g,t}$] generation of PSH unit $g$ at time $t$ in day-ahead solution [MW];
	\item[$Q^{pump,DA}_{g,t}$] pumping of PSH unit $g$ at time $t$ in day-ahead solution [MW];
	\item[$dT$] interval length [one hour].
\end{IEEEdescription}

Variables [units]:

\begin{IEEEdescription}[\IEEEusemathlabelsep\IEEEsetlabelwidth{$V_1,V_2.5$}]
	\setlength{\itemsep}{2pt}
	\item[$e_{r,t}$] continuous variable, energy stored in the reservoir $r$ at time $t$ [MWh];
	\item[$e_{r,s,t}$] continuous variable, energy stored in the reservoir $r$ in scenario $s$ at time $t$ [MWh];
	\item[$u_{g,t}$] binary variable, commitment variable of unit $g$ during time interval $t$ [NA];
	\item[$p_{g,t}$] continuous variable, amount of generation at unit $g$ during time interval $t$ [MW];
	\item[$q_{g,t}^{gen}$] continuous variable, amount of generation at a PSHU $g$ during time interval $t$ [MW];
	\item[$q_{g,t}^{pump}$] continuous variable, amount of pumping load at a PSHU $g$ during time interval $t$ [MW];
	\item[$q_{g,s,t}^{gen}$] continuous variable, amount of generation at a PSHU $g$ in scenario $s$ during time interval $t$ [MW];
	\item[$q_{g,s,t}^{pump}$] continuous variable, amount of pumping load at a PSHU $g$ in scenario $s$ during time interval $t$ [MW];

\end{IEEEdescription} 

%Derived Data [units]:

%\begin{IEEEdescription}[\IEEEusemathlabelsep\IEEEsetlabelwidth{$V_1,V_2.5$}]
	\setlength{\itemsep}{2pt}
	%\item[$H_{gt}^{k,h}$] the piecewise cost of the $k^{th}$ segment of piecewise approximation of the production cost or bid price of configuration $h$ at generator $g$ at period $t$  [\$/MW];
	%\item[$P_{gt}^{k,h}$] the break point of the $k^{th}$ segment of piecewise approximation of the production cost of configuration $h$ at generator $g$ at period $t$ [MW].
	%\item[$C_{g,t}^{s}$] the bid pumping price of unit $g$ at time $t$ [\$/MW]. 
	
%\end{IEEEdescription}

Auxiliary Variables [units]:

\begin{IEEEdescription}[\IEEEusemathlabelsep\IEEEsetlabelwidth{$V_1,V_2.5$}]
	\setlength{\itemsep}{2pt}
	\item[$C(q,u)$] dispatch and commitment cost function of a generating unit [\$];
	\item[$W_r$] robust auxiliary variable [\$].
\end{IEEEdescription}

\section{Introduction}
\subsection{Background and Motivation}
\IEEEPARstart{P}{}umped storage hydro (PSH) can mitigate the increasing variations and uncertainties in a modern power system. A PSH unit (PSHU) not only has a large power capacities, but also can switch between modes and ramp very fast. In addition, a PSHU normally has a large storage capacity and can continuously charge and discharge in a long duration in comparison to the other available electric storage technologies. The combination of all of these characteristics makes a PSHU capable of providing a wide range of services to the electric grid such as weekly and daily smoothing of loads, spinning reserve and voltage/frequency control \cite{barton2004energy}. What is more, the PSHU has been integrated and operated in the power system for decades with valuable operational experiences. There are 22.9 GW of PSH generating capacity in the US \cite{eiamost2018}. Due to the increasing integration of intermittent renewable resources and the growing active demand responses, more uncertainties and variations are introduced to the modern power system. PSHUs can be used to absorb variations and help system operators to mitigate some of the system uncertainties. Therefore, it can be critical to efficiently use the PSHUs and leverage their flexibility in the operation of the power systems. 

%However, the market model for a PSHU were also developed decades ago and they are still in use in the current system that is experiencing a dramatic portfolio changes now. Some of these market models are outdated and they can be improved to unlock more flexibility from the PSHUs. The first step is to update the Day-ahead (DA) market model and incorporate the PSHUs in an independent system operator's (ISO) unit commitment and economic dispatch (UCED) problem. A PSHU model with case studies in the MISO DA market is proposed in \cite{huang2021configuration}.
% The key is to represent the state-of-charge (SOC) of the PSHUs in the UCED optimization and allow the ISO to optimize the charging and discharging decisions. PJM is the only ISO in the US that currently implements the SOC constraints of a PSHU in their DA optimization model.     

Due to the increasing uncertainties partially driven by the integration of renewable energy resources, there are increasing opportunities for storage units to actively participate in the real-time market operation. However, in the current Mid-continent Independent System Operator (MISO) system, PSH participants try to stick to their day-ahead (DA) plan in the real-time (RT) market to avoid the risk of volatile prices in the RT market. DA and RT arbitrage revenue have been studied for Pennsylvania-New Jersey-Maryland Interconnection (PJM) and California Independent System Operator (CAISO) in \cite{salles2017potential} and \cite{byrne2018opportunities} respectively. Both studies conclude that it could be more beneficial to have storage unit actively participating in both DA and RT markets. A study in multiple US markets shows that PSH is one of the storage technologies that have the greatest potential for DA and RT market arbitrage \cite{bradbury2014economic}. 

A rolling three hours look-ahead commitment (LAC) is applied in MISO system to assist operators to make commitment decisions in the RT market. Based on the previous work on the PSH DA model, this paper proposes two PSH models in the LAC to explore the potential of having the PSHUs optimized in MISO real-time (RT) market clearing software. It is assumed that the PSHUs participate in the DA market, establishing a DA financial position, and also participate in the RT market and can adjust the positions in the first interval of each of the LAC optimizations. The challenge for including PSH in a LAC formulation is two folds. First, the three hour LAC window is too short for the the charging and discharging cycle of a PSHU. Second, the real-time market operation requires the security constrained unit commitment (SCUC) to be solved in a few minutes. The key issues are the incorporation of the real-time uncertainties beyond the LAC window, the appropriate representation of the value of the stored energy and the efficient stochastic modeling that does not impose major computational burden.

\subsection{Literature Review}

The research in the hydro/pumped storage hydro in the multistage market has been active globally. The previous research works in the related area can be summarized in two categories. The first group of work solve the profit maximization in the RT market or in the joint DA and RT markets and look for the optimal bidding strategy from the storage market participant stand point. The second group of work solve the unit commitment and economic dispatch for the system and leverage the storage units through the optimization.    

Different approaches have been used in the first category where the profit maximization is solved for the storage unit. The DA and RT markets are considered and jointly solved in several works. A bi-level optimization is presented in \cite{wang2017look} where the RT market is formulated as the lower-level problem. The conditional value at risk (CVaR) is applied to manage the uncertainty. A stochastic optimization model is proposed in \cite{krishnamurthy2017energy} that shows the significant improvement of stochastic approach benchmarked with the deterministic benchmark. In \cite{berrada2016valuation}, a linear programming approach is proposed to dispatch a solar-storage unit in DA and RT market. The storage unit is used in the RT market to respond to the solar and load forecast error from DA. The co-optimization of energy and regulation reserve as a price taker is studied for variable speed pumped storage hydropower plants with the Iberian power system \cite{chazarra2017optimal}. Bidding strategy in sequential electricity markets has been studied with the Nordic system \cite{boomsma2014bidding}.

A RT rolling optimization is used in the studies of the operation of the storage unit in the RT market. In \cite{khatami2019look}, a RT rolling look-ahead optimization model is proposed for Compressed Air Energy Storage (CAES). The unit commitment decision made for the CAES in the DA market is enforced in the RT market. The RT optimization of a storage in the Ontario ISO is studied in \cite{khani2014real}. The price forecast is updated only in a three hour rolling look-ahead window and the price forecast for the rest of the day is based on the DA price forecast. In \cite{bhatti2020combined}, DA and RT rolling profit maximization is investigated. This work investigated optimal reactive power dispatch in a distribution system. In \cite{liu2014dispatch}, based on the realized actual Location Marginal Price (LMP) in RT market, the pumped storage hydro unit is applied with a policy function to react to the wind forecast error.
%a LMP forecast is used to optimize the profit of a wind and a pumped storage hydro unit in the DA market. 

A dynamic programming approach is used to study the wind-storage hybrid model in the RT market dispatch in \cite{nguyen2012new}. A deterministic LMP forecast is used in the RT simulation.

Several different approaches have been used in the second group of studies to leverage the storage units in the RT unit commitment and economic dispatch problem. In \cite{o2017efficient}, multi-stage deterministic simulation are used to study the impact of different combination of reserve products provided in the system on the system costs and the profit of PSHUs. 

%This work focused on the various of reserve products and the pumped storage hydro units' participation in a system. The impact of different combination of reserve products provided in the system on the system costs and the profit of pumped storage units is explored. However, the uncertainty for a storage unit in a RT operation is not explicitly addressed. 

Stochastic programming model is used to incorporate uncertainties in the operation of storage units in a RT or close to RT market. The reserve capacity procurement has been studied for a hydropower scheduling problem with Norwegian watercourse \cite{naversen2020accounting}. A large-scale stochastic programming model that is used for weekly hydrothermal dispatch and spot pricing of the Brazilian power system is presented in \cite{diniz2018short}. In \cite{bakirtzis2018storage}, the operation of storage devices is studied in a rolling stochastic unit commitment model using both single point and probabilistic forecast on wind and load. In \cite{papavasiliou2017application}, a multistage stochastic programming formulation is proposed to optimize storage dispatch in RT operations.

Policy function has been used to address this challenge. In \cite{li2015flexible}, the operation range of a battery storage in a Look-Ahead Commitment (LAC) is determined by the stochastic DA unit commitment (UC) solution that is closest to the realized renewable scenario. Generic multi-stage recourse policies are investigated for the rolling unit commitment and dispatch problem in \cite{warrington2015rolling}. An affine dispatch policy for renewable and storage units is studied in the multi-stage unit commitment problem with the Polish system \cite{lorca2016multistage}.

\subsection{Contributions}

There are two research questions this paper tries to answer. The first question is how uncertainties can be efficiently incorporated in optimizing PSHU in a RT LAC rolling window without introducing major computational burdens. The existing literature in the unit commitment and economic dispatch (UCED) problem point to the stochastic programming direction \cite{bakirtzis2018storage}, \cite{papavasiliou2017application}, \cite{warrington2015rolling}, \cite{li2015flexible}. However, it is computationally very expensive to incorporate all the uncertainties from load and renewable resources and take the input scenarios in a LAC especially for a large system like MISO. The second question is how to address the risk of the uncertainties when dispatching a PSH in the RT market. While many of the previous work on storage optimization consider uncertainties \cite{krishnamurthy2017energy}, \cite{liu2014dispatch} and \cite{berrada2016valuation}, only a few address the risk \cite{wang2017look} and none of them considers the risk management from the system stand point.

Although price forecast is used in the study of storage profit maximization \cite{khani2014real}, \cite{liu2014dispatch}, \cite{nguyen2012new}, a probabilistic scenario based price forecast hasn't been used in a SCUC problem from the system point of view.

The contribution of this paper is summarized below:

\begin{itemize}
	\item A novel price forecast based stochastic model is proposed for the PSHU LAC optimization. In the proposed formulation, only PSHU is included in the periods after the LAC which is much simpler and computationally efficient than including the full system model for the entire day in each LAC.
	\item An easier to implement scenario-based price forecast is used in the proposed PSHU LAC models.
	\item A novel formulation of robust model that reflects the actual risk-averseness of PSH owner is proposed for the PSH LAC optimization. It is demonstrated that market efficiency can be improved even with the risk-averse model.
	\item The proposed models are prototyped in a MISO system and benchmarked with three other models including the current practice. The case studies provide realistic references.
\end{itemize}

\section{LAC Formulations for PSH}
In this section, we propose two models that use probabilistic price forecasts in the LAC formulation to capture the real-time market uncertainty beyond the end of the LAC window. It is assumed the unit commitment decision of a PSHU can be changed in the LAC with no impacts from the DA solution. The DA commitments for other resources are respected in LAC. The DA dispatch solution for a PSHU is used only in the robust model as a reference point as is discussed in section \ref{sec:RobModel}. Notice that the scope of the study is the single day LAC and RT market. Therefore, both models are focused on the daily operation of the PSHU in the LAC.
%\ref{fig:state_gram}

\begin{figure}[!h]
	\centering
	\includegraphics[width=0.42\textwidth]{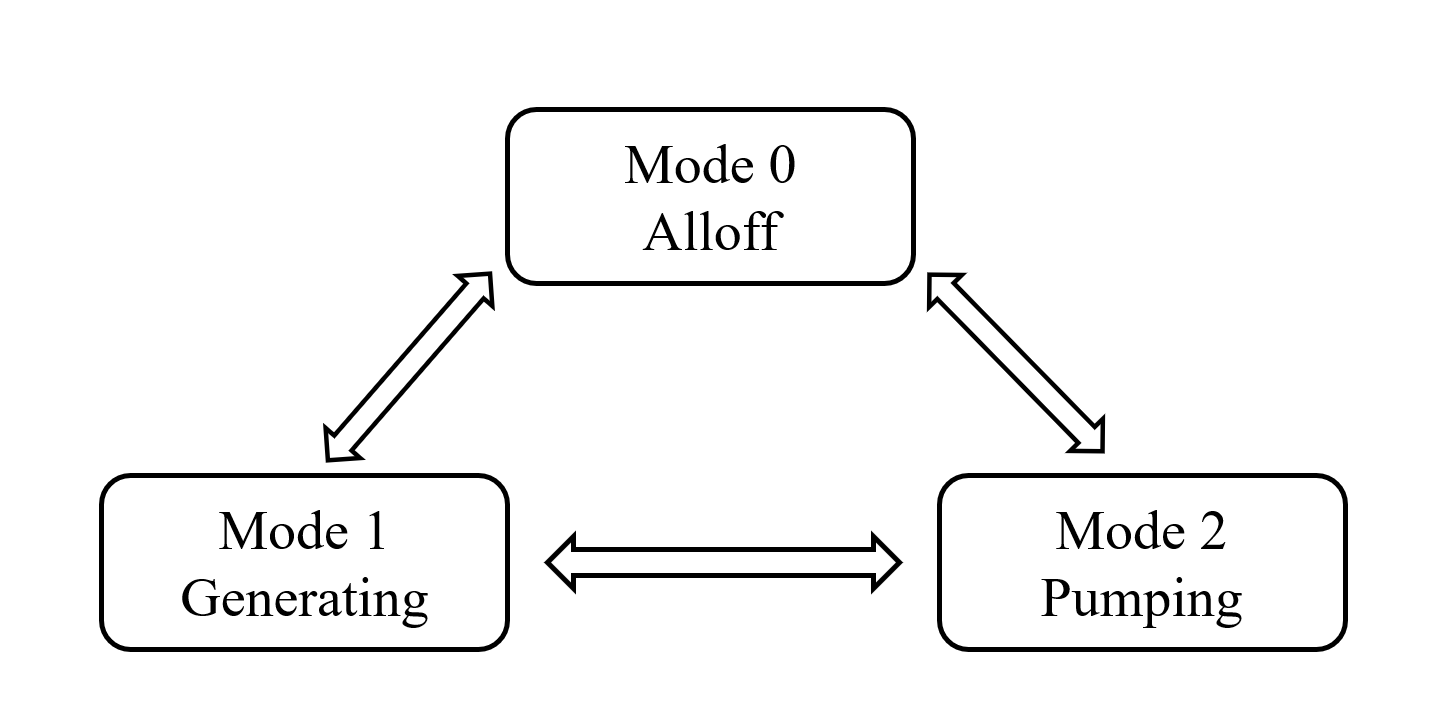}
	\caption{Mode transition diagram of a PSHU in two consecutive periods \cite{huang2021configuration}.}
	\label{fig:state_gram}
\end{figure}

The LMP forecast is used to provide guidance to the PSHU in a LAC. LAC in MISO contains system status and short term load forecast for the next three hours. However, that is not enough for a PSHU which typically can operate a daily cycle. 
%It can be quite challenging to extend the length of LAC windows due to the tight computational limit in the real-time operation. In addition, since the focus of the paper is on the PSHU dispatch, it will facilitate a potential implementation if the current LAC formulation for the system is kept unchanged. Therefore, for the operation of the PSHU, it is important to effectively manage the units' \textcolor{blue}{state of charge (SOC)} at the end of each of the series of LAC windows to determine how much energy is left in the reservoir to be used in the future intervals of the day.
The key is to find a good way to reflect the system information from the future (after the LAC window) to the present (inside the current LAC window) so that the LAC could optimize the SOC of the PSHU with the rest of the system while being cognizant of the conditions in the future intervals. Therefore, we propose to use LMP forecasts in the intervals post to a LAC window to optimize a PSHU. The LMP forecast methodology is discussed in detail in section \ref{sec:price forecast}. In this section, we assume a probabilistic LMP forecast is available.

%One approach is to dualize the intertemporal SOC constraints and include the shadow price of the SOC constraints at the boundary of the LAC window to the objective function. This approach would give an accurate value of the SOC at the end of a LAC. However, there is no historical data on the shadow price of the SOC constraints. So, it is practically not feasible to forecast the SOC shadow price in the LAC optimization. 

The LAC formulations for a PSH is developed based on a configuration-based modeling of PSHU that represents all feasible operation modes and the state-of-charge (SOC) of a PSHU as described in \cite{huang2021configuration}. A pumped storage hydro plant can contain multiple units and each of them will be modeled individually; however, there are only three operation modes in a PSHU, namely generating, pumping, and offline, which are mutually exclusive as shown in Fig \ref{fig:state_gram}. Transitions are allowed between each pair of these modes indicated as the double-headed arrows. %\textcolor{blue}{Because the focus of this paper is on the incorporation of uncertainty and risk management in the LAC and RT optimization, we have accepted some simplifications in the PSHU model in this paper. For example, a constant efficiency is provided by the PSHU owners. Also, the water to energy transformation is provided by the PSHU.} \textcolor{blue}{The details of the PSHU unit commitment model can be found in the Appendix \eqref{eq:mutual}-\eqref{eq:hydrobox}.} 

\subsection{Stochastic PSH Model}\label{sec:StoModel}
\begin{figure}[!h]
	\centering
	\includegraphics[width=0.42\textwidth]{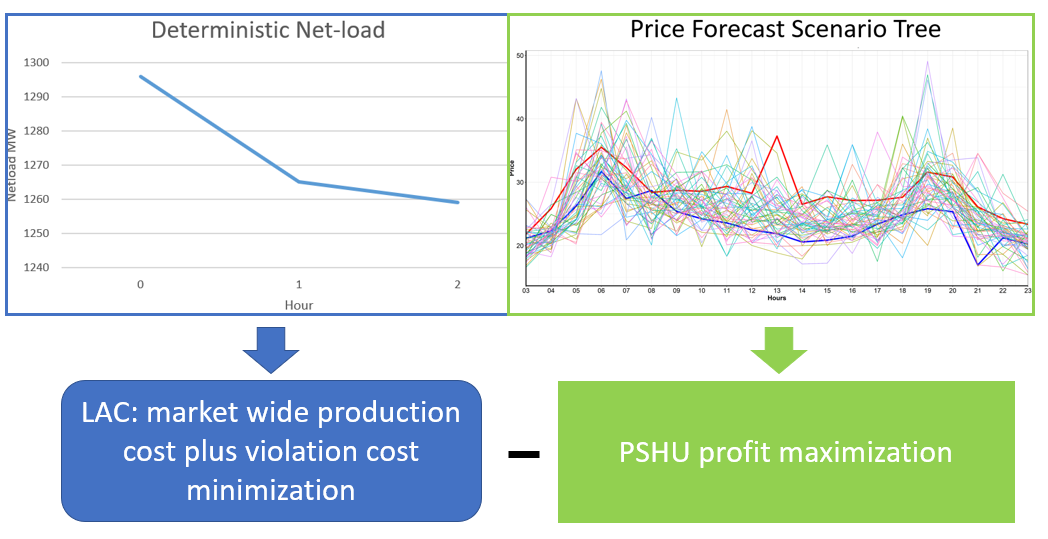}
	\caption{Risk-neutral Stochastic PSH LAC Illustration}
	\label{fig:Two_stage_illustration}
\end{figure}

A two-stage stochastic PSH Model in LAC is proposed in this subsection. The first stage decisions are the unit commitment and dispatch decisions in a LAC problem. The second stage decisions are the PSHU commitment and dispatch decisions in the future intervals that starts from the first interval after a LAC until the last interval in the day. The commitment and dispatch of the PSHU in the intervals after the LAC is optimized using the LMP forecast. The revenue of the PSHU after the LAC window are subtracted in the objective as shown in the second term in \eqref{eq:objective}. As illustrated in Fig \ref{fig:Two_stage_illustration}, this formulation combines 1) market wide production cost plus violation cost minimization within the LAC window that take a deterministic net-load as input and 2) PSHU profit maximization after the LAC window that takes price forecast scenarios as input. Notice that, for each LAC window, the problem is formulated and solved as a single Mixed-Integer linear programming (MILP) model.

\subsubsection{Objective Function}
The formulation of the stochastic PSH model in LAC is presented in \eqref{eq:objective}. 

\begin{equation} 
\begin{split}
&\underset{p,q,u}{\min}[\sum_{g\in \mathcal{G}} \sum_{t=t_1}^{t_{end}}C(p_{g,t},q_{g,t},u_{g,t}) \\& - \sum_{s \in \mathcal{S}}\sum_{t=t_{end}+1}^{T}\sum_{g\in \mathcal{G}_{psh}} P_s LMP_{g,s,t}^{t_0} (q^{gen}_{g,s,t}-q^{pump}_{g,s,t})]. 
\end{split}
\label{eq:objective}
\end{equation}

The first term in \eqref{eq:objective} is the objective function for a LAC problem. The production cost $C (p_{g,t},q_{g,t},u_{g,t})$ is minimized in a LAC window in intervals that start at $t_1$ and end at $t_{end}$. %\textcolor{blue}{The production cost includes unit generation costs, generation no load costs, start up/shut down costs, different reserve costs and the penalty on the transmission constraints violation. A PSHU does not have generation costs, but it has start-up and shut-down costs for the unit commitment decisions.} 
It is assumed, except for start-up costs, the operation and maintenance cost is negligible for a PSHU. The second term without the negative sign reflects the expected revenue from dispatching the PSHU under the forecasted pricing scenarios in the future intervals post to LAC. Different PSHU generation and pumping values are allowed in each scenario. It is acknowledged that, strictly speaking, causality is violated by the implicit assumption that the generation and pumping values can be chosen for all intervals in a given scenario. With the weighted probability $P_s$, the probabilistic LMP forecast $LMP_{g,s,t}^{t_0}$ is provided for each interval after the LAC and the forecast is updated at $t_0$ that is one interval before the start of each LAC window $t_1$. 

\subsubsection{Power Balance Constraints}
\begin{equation} 
\begin{split}
\sum_{g\in \mathcal{G}} p_{g,t} + \sum_{g \in \mathcal{G}_{psh}}q_{g,t}^{gen} =D_t+\sum_{g \in \mathcal{G}_{psh}}q_{g,t}^{pump}, \forall t \in [t_1,t_{end}],
\end{split}
\label{eq:powerbalance1}
\end{equation}

%\begin{equation} 
%\begin{split}
%\sum_{g\in \mathcal{G}} q_{g,t}  =D^{DA}_t, \forall t \in [t_{end}+1,T]. 
%\end{split}
%\label{eq:powerbalance2}
%\end{equation}

The PSHU is fully optimized within the LAC window given a deterministic forecast of the demand within the LAC window. In the power balance constraint within the LAC window $\forall t \in [t_1,t_{end}]$, the deterministic generation of the PSHU, $q_{g,t}^{gen}$, is included on the left hand side of power balance constraint \eqref{eq:powerbalance1} and the deterministic pumping load of the PSHU, $q_{g,t}^{pump}$, is considered as demand on the right hand side of the power balance constraint \eqref{eq:powerbalance1}.

\subsubsection{The Private Constraints for a PSHU in the LAC}

The private constraints of a PSHU model, such as the state transition constraints, SOC constraints and mutually exclusive constraints, are the same as the DA model described in (3)-(11),(13),(14) in \cite{huang2021configuration}. The PSHU private constraints are modeled in the intervals from the start of the LAC window $t_1$ until the end of the operating day $T$. The private constraints for a PSHU model are deterministic within the LAC intervals and they are defined for each scenario in the intervals after the LAC. In the following, the SOC constraints are explained in detail as an example. Notice that the time interval is taken as hourly in this paper, therefore hourly time interval $dT$ is timed with the generation and pumping capacity in equations \eqref{eq:StorageEnergyBal1}-\eqref{eq:StorageEnergyBal3}. The rest of the private constraints are formulated similarly. The detailed description of each of the constraints can be found in (3)-(11),(13),(14) in \cite{huang2021configuration}.

\begin{equation}
\begin{split}
e_{r,t+1}=e_{r,t} + \sum_{g\in \mathcal{G}_{psh,r}} \eta_g^{pump}q_{g,t}^{pump} dT\\-\sum_{g\in \mathcal{G}_{psh,r}} \frac{q_{g,t}^{gen}}{\eta_g^{gen}} dT,  \forall r \in \mathcal{R}, \ \ \forall t \in [t_1,t_{end}-1],
\end{split}
\label{eq:StorageEnergyBal1} 
\end{equation}
\begin{equation}
\begin{split}
e_{r,s,t_{end}+1}=e_{r,t_{end}} + \sum_{g\in \mathcal{G}_{psh,r}} \eta_g^{pump}q_{g,t_{end}}^{pump}dT\\ -\sum_{g\in \mathcal{G}_{psh,r}} \frac{q_{g,t_{end}}^{gen}}{\eta_g^{gen}}dT,  \forall r \in \mathcal{R}, \ \ \forall s \in \mathcal{S},
\end{split}
\label{eq:StorageEnergyBal2} 
\end{equation}
\begin{equation}
\begin{split}
e_{r,s,t+1}=e_{r,s,t} + \sum_{g\in \mathcal{G}_{psh,r}} \eta_g^{pump}q_{g,s,t}^{pump}dT \\ -\sum_{g\in \mathcal{G}_{psh,r}} \frac{q_{g,s,t}^{gen}}{\eta_g^{gen}}dT,   \forall r \in \mathcal{R}, \ \ \forall s \in \mathcal{S}, \ \ \forall t \in [t_{end}+1,T].
\end{split}
\label{eq:StorageEnergyBal3} 
\end{equation}

In the intervals within the LAC window, the deterministic SOC constraints are formulated in \eqref{eq:StorageEnergyBal1}. The energy stored in the PSH system is linked between every consecutive time interval. Notice that there can be more than one PSHU sharing a reservoir in the model. Parameters $\eta_g^{gen}$ and $\eta_g^{pump}$ are the efficiencies of generating and pumping modes that indicates the energy loss on both modes. In the intervals after the LAC window, starting at $t_{end}+1$ until the end of the operating day $T$, the SOC constraints are formulated for each scenario $s$ in \eqref{eq:StorageEnergyBal3}. For the inter-temporal SOC constraint, we need to specifically address the constraint when it crosses between the interval within a LAC window and the interval after the LAC window. In \eqref{eq:StorageEnergyBal2}, the SOC changes from the last interval of the LAC, $t_{end}$, and the first interval after the LAC, $t_{end}+1$, are defined for each scenario $s$. Notice that the SOC variable and generation and pumping variables at the last interval of LAC namely $e_{r,t_{end}},q_{g,t_{end}}^{gen},q_{g,t_{end}}^{pump}$ are deterministic and the SOC in the first interval after LAC is defined for each scenario, $e_{r,s,t_{end}}$. Therefore, by \eqref{eq:StorageEnergyBal2}, every scenario based SOC variable in the intervals after LAC is linked to the last deterministic SOC variable within the LAC.

\begin{equation} e_{r,t_1}=E_{r,t_1}, \ \ \forall r \in \mathcal{R},
\label{eq:socb}
\end{equation}
\begin{equation} e_{r,s,T+1}=E_{r,T+1}, \ \ \forall r \in \mathcal{R} \ \ \forall s \in \mathcal{S},
\label{eq:soce}
\end{equation}
\begin{equation} \underline{E_r} \leq e_{r,t} \leq \overline{E_r}, \ \ \forall r \in \mathcal{R}, \ \ \forall t \in [t_1,t_{end}],
\label{eq:socbox}
\end{equation}
\begin{equation} \underline{E_r} \leq e_{r,s,t} \leq \overline{E_r}, \ \ \forall r \in \mathcal{R},\ \ \forall s \in \mathcal{S}, \ \ \forall t \in [t_{end}+1,T].
\label{eq:socboxs}
\end{equation}

The initial SOC in the LAC is given by the previous SOC solution indicated as $E_{r,t_1}$ in \eqref{eq:socb}. The SOC starting point in the first LAC is given by the PSHU and it is the same as the starting SOC in the DA problem. The SOC variable at the end of the day, $e_{r,s,T+1}$, is fixed to the given target $E_{r,T+1}$, that is the SOC at the end of the day in the DA solution, in each scenario in \eqref{eq:soce}. The end of the day SOC target is calculated from the historical production data, equation \eqref{eq:soce}, so as to make a fair comparison between the proposed models and the current practice. The upper and lower limit are enforced on deterministic SOC variables within LAC and scenario based SOC variables post to LAC for each scenario in \eqref{eq:socbox} and \eqref{eq:socboxs} respectively.

\subsection{Robust PSH Model}\label{sec:RobModel}

Current PSH owners are usually concerned about the exposure to uncertain RT price. Therefore, they typically stay with the DA positions. To address this risk-averse concern in the existing practice, a robust risk-management formulation is developed and it can be applied to the stochastic PSHU LAC formulation described in section \ref{sec:StoModel}.

\subsubsection{Objective Function and Risk Management Constraint}
The formulation of the robust PSH model in LAC is presented in \eqref{eq:Robobjective} and \eqref{eq:riskmanage}. 

\begin{equation} 
\begin{split}
\underset{p,q,u,v}{\min}(\sum_{g\in \mathcal{G}} \sum_{t=t_1}^{t_{end}}C(p_{g,t},q_{g,t},u_{g,t}) + \sum_{r \in \mathcal{R}}W_r),
\end{split}
\label{eq:Robobjective}
\end{equation}

\begin{equation}
\begin{split}
W_r \geq &- \sum_{t=t_{end}+1}^{T} \sum_{g \in \mathcal{G}_{psh,r}}LMP_{g,s,t}^{t_0}[(q_{g,s,t}^{gen}-q_{g,s,t}^{pump})- \\&(Q_{g,t}^{gen,DA}-Q_{g,t}^{pump,DA})], \ \ \forall r \in \mathcal{R}, \forall s \in \mathcal{S}. 
\end{split}
\label{eq:riskmanage}
\end{equation}

In the risk-management formulation, the objective is updated in \eqref{eq:Robobjective}. Notice that the first term is the system production costs and that is the same to the first term in the stochastic objective in \eqref{eq:objective}. The difference is in the second term of the objective function. In \eqref{eq:objective}, the cost/negative profit of the PSHU in the intervals after the LAC is weighted by the probability of each scenario. Therefore, the model presented in \eqref{eq:objective} is a risk-neutral formulation. However, in \eqref{eq:Robobjective}, the cost of each PSH plant $r$ is represented by an auxiliary variable $W_r$ which represents the worst case scenario that is defined in \eqref{eq:riskmanage}. The right-hand side of \eqref{eq:riskmanage} is the negative profit of the PSHU in the RT market after the LAC intervals in each scenario. The RT profit is calculated as RT LMP forecast at each scenario $LMP_{g,s,t}^{t_0}$ times with gen/pump difference between its solution in RT market in a scenario $(q_{g,s,t}^{gen}-q_{g,s,t}^{pump})$ and the solution in the DA market $(Q_{g,t}^{gen,DA}-Q_{g,t}^{pump,DA})$. Constraint \eqref{eq:riskmanage} limits each cost variable $W_r$ to be the cumulative cost of the PSHU from the first interval after LAC, $t_{end}+1$, to the end of the day, $T$, in the worst-case scenario (that is the largest cost to the system or the lowest RT profits to the PSHU) based on the probabilistic LMP forecast. Therefore, since the worst-case PSHU cost is minimized in the objective, it is a robust or risk averse formulation. The rest of the stochastic PSHU model remains unchanged from \ref{sec:StoModel}.
With the proposed risk-management formulation in \eqref{eq:Robobjective} and \eqref{eq:riskmanage}, the solution for a PSHU will only deviate from the DA solution if it is profitable in every post-LAC price scenario. Therefore, \eqref{eq:Robobjective} and \eqref{eq:riskmanage} address the industry concern of financial loss in the RT market. As in Section \ref{sec:StoModel}, it is acknowledged that, strictly speaking, causality is violated by the implicit assumption that the generation and pumping values can be chosen for all intervals in a given scenario.  
It is noted that equation \eqref{eq:soce} is enforced so as to make a fair comparison between the proposed models and the current practice. In a production setting, this constraint can be relaxed and thereby would prevent the robust model being overly strict.

\section{Price Forecast Methodology}\label{sec:price forecast}
RT LMP is challenging to forecast due to its volatility that is caused by factors such as generation availability and uncertainty, fuel prices, load forecast uncertainty, weather condition, and market participant unpredictable behavior, and that makes RT price forecasting a challenging problem \cite{singhal2011electricity}. This has motivated numerous efforts to provide more information about the uncertainty associated with the single point price forecast rather than just a simple statistical summary. A general seasonal periodic regression model with ARIMA and Fractional ARIMA (FARIMA) is considered in \cite{Koopman2007}.  For model identification and estimating optimum parameters for Auto-Regressive (AR) and Moving Average (MA) terms, partial autocorrelation function (PACF) and Autocorrelation function (ACF) is used in all ARIMA family time series models. In the context of electricity market, for the purpose of energy system planning and operations, the need for probabilistic electricity price forecast (EPF) becomes more prominent. Reference \cite{Nowotarsk2018} provides a thorough review of probabilistic electricity price forecasting (EPF). The main approaches discussed to deal with probabilistic EPF includes creating Prediction Intervals \cite{Werona2008}, \cite{Nowotarski2015},  distribution based probabilistic forecasts  \cite{Smith2008} \cite{Maciejowskaab2016}, boot-strapped PIs, which is commonly used in neural network EPF studies \cite{Clementsa2007}, and Quantile Regression Averaging (QRA) which combines multiple point forecasts of individual time series models with the concept of quantile regression \cite{Katarzyna2016}

In this paper we first use a single point forecast namely ARIMAX model for RT-LMP. Based on the single point forecast, a probabilistic forecast with statistical scenarios is proposed. The method used for point forecasting is Auto Regressive Integrated Moving Average with Exogenous variables ARIMAX, where the idea of using covariates $X$ into the forecasting model is for increasing the accuracy of forecasting the variable of interest. Since a single quantile forecast does not provide enough information for most optimization and decision-making processes, there is a potential economic benefit from good estimates of the uncertainty associated with RT price forecasts. The non-parametric probabilistic forecasts is used to cover the uncertainty range in the prediction of RT-LMP. However, the general form of probabilistic LMP forecasts do not reflect the interdependence structure of errors coming from previous time horizons. The non parametric forecasts using a state-of-the-art statistical scenario generation methodology that considers the interdependence structure of errors used for wind production \cite{Pinson2008} is adopted.

\subsection{ ARIMAX-  Point Forecast Methodology for RT-LMP}\label{sec:ARIMAX}
ARIMAX is a type of linear regression model that uses past observation of the target variable (RT-LMP) along with some covariates called exogenous variables (Day-ahead-LMP) to forecast future values. In this model, Auto-Regressive (AR) means the currently observed value of target variable is some linear combinations of its past values. Moving Average (MA) uses the past errors to forecast the target variable. Integrated, means that the future changes in the target variable is linear function of its past change, which are evaluated by applying a differencing step to data. Given time series data (RT-LMP) and exogeneous data (Day-ahead-LMP), represented by the variable $x_(m,t)$, where $p$ is the number of auto-regressive lags, $d$ is the degree of differencing, and $q$ is the number of moving average lags, the point forecasting model is written as:
\begin{equation} 
	y_t=\sum_{i=1}^{p} \phi{_i}  y_{t-i}+\sum_{j=1}^{q} \theta_{j}  \epsilon_{t-j}+ \sum_{m=1}^M \beta_{m}  x_{m,t}+\epsilon_{t},\hspace{4pt} \epsilon \sim N(0,\sigma^2 ),  
	\label{eq:ARIMAX_model} 
\end{equation}                     
Mean Absolute Percentage Error (MAPE) is used to evaluate the performance of the ARIMAX model vs ARIMA and Seasonal ARIMA (SARIMAX). Considering one whole year data as our test set, in 95\% of test days ARIMAX outperformed ARIMA and in 87\% of test days ARIMAX surpassed the performance of Seasonal-ARIMAX (SARIMAX) by reporting lower error number using MAPE as the performance measurement. For the scenario based probabilistic forecast, the quality is assessed by how effective the forecast scenarios could guide the PSHU in the RT market dispatch. Therefore, a profit maximization simulation is used to assess the quality in comparison to the after the fact LMP in \cite{ghesmati2022probabilistic}. The average ratio of profit gained from the price forecast scenarios to the profit using the after the fact LMP is $13.69\%$ with standard deviation of $25.4$. In addition, the comparison of the results between the stochastic model and the perfect LAC model in Table I and Table II in the manuscript can also be served as a quality assessment for the price forecast scenarios.

\subsection{Probabilistic Forecast in the form of Statistical Scenarios for RT-LMP}\label{sec:probabilistic forecast}
\begin{figure}[!h]
	\centering
	\includegraphics[width=0.42\textwidth]{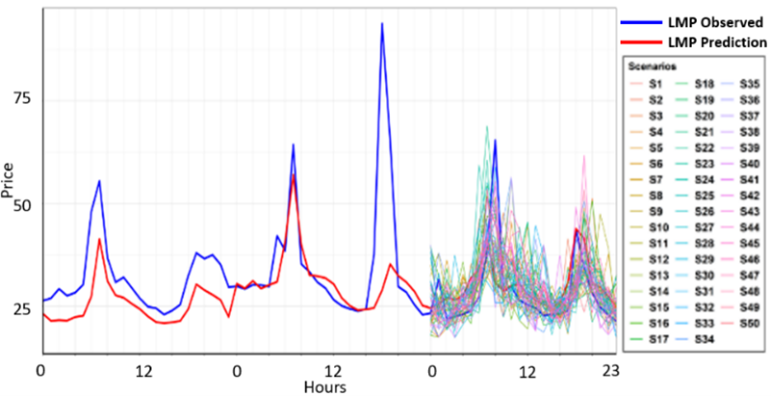}
	\caption{RT-LMP point forecast and its associated statistical scenarios; The scenarios show the expected range of forecasting uncertainty.}
	\label{fig:Price_Forecast}
\end{figure}

For most optimization operations and decision-making processes, a single quantile forecast as shown by the thick red line in Fig. \ref{fig:Price_Forecast} is not sufficient for making an optimal decision for a given time horizon. Assuming $p_{(t+k)}$  is the RT-LMP forecast at time $t+k$. If we do not have a certain assumption on the shape of distribution function then we can use the results of a quantile regression model of price data to provide a forecast for Probability Density Function (PDF) shown in $f_{t+k}$ for any look-ahead time $t+k$. Therefore, to reconstruct the conditional distributions for the time series price values at any given time and forecast the Probability Density Function (PDF) we can directly use the quantiles out of fitting the Quantile-Regression model as follows:
\begin{equation} 
\hat{f}_{t+k} = \{\hat{q}^ {\alpha_{i}}_{(t+k)} \hspace{2pt} 0\leq \alpha_{1} \leq \cdots \leq \alpha_{m} \leq 1, \hspace{3pt}, 0\leq m \leq 1.\}
	\label{eq:pdf_equ} 
\end{equation} 
where $\hat{q}_{(t+k)}$ are the quantile functions. then the random variable $W_k$ whose realization $W_k^t$  at time $t$ is defined by
\begin{equation} 
	W_k^t=\hat{F}_{t+k} (p_{t+k}) \hspace{3pt}  ,  \forall t
	\label{eq:w_equ} 
\end{equation} 
 is uniformly distributed on the unit interval $U[0,1]$. In order to transform the variable $W_k^t$ with uniform distribution to a random variable $X_k^t$ with standard normal distribution we use the probit function to apply the transformation: $X_k^t = \Phi ^{-1} (W_k^t), \forall t$, where $\Phi^{-1}$ is the inverse of the Gaussian cumulative distribution function or probit function.\\
For any look-ahead time, the vector $X$ as a transformed random vector of forecasted price in 24-hours look ahead horizon
\begin{equation} 
X= (X_1, X_2, \cdots, X_{24})^T \sim N(\mu_0,\Sigma ) 
	\label{eq:vectorX} 
\end{equation} 
follows a multivariate Gaussian distribution where $\Sigma$ is Covariance matrix and is defined recursively. To estimate the sample covariance Matrix of the forecast error we follow \cite{Pinson2008}  and use a recursive algorithm, where $\lambda \in [0, 1)$ is the forgetting factor, and the covariance matrix is initialized by setting it equal to the Identity matrix.

\begin{equation} 
	\sum_t=\lambda \sum_{t-1} + (1-\lambda) X^t {X^t}^T
	\label{eq:recursive} 
\end{equation} 

%Assuming $p_{t+k}$  is the RT-LMP forecast at time $t+k$, then the random variable $W_k$ whose realization $W_k^t$  at time t is defined by $W_k^t=\hat{F}_{t+k} (p_{t+k} ),  \forall t;$ is uniformly distributed on the unit interval $U[0,1]$. 
%	The next step to create random variables with Gaussian distribution uses the quantile function of the standard Normal distribution, known as the Probit function, as follows:

%\begin{equation} 
%	X_k^{t}= \Phi^{-1} (W_k^t), \hspace{2pt} \forall t
%	\label{eq:probitFunc} 
%\end{equation} 	
%With this transformation we are able to project the forecasted price variable $p_{t+k}$ into standard Normal space shown as variable $X_k \sim N(0,1)$.  Now The next step is capturing the interdependence structure of forecasted errors in look ahead time horizon, 24 hours in our case.
%	\subsection{Creating Covariance Matrix} \label{sec:con_matrix}
%	$X={(X_1, X_2,…, X_{24})}^ T$ is the transformed random vector of forecasted price in 24-hours look ahead horizon, which by definition has Gaussian distribution $X\sim N(\mu_0,\Sigma)$. 
  
\begin{figure}[!h]
	\centering
	\includegraphics[width=0.42\textwidth]{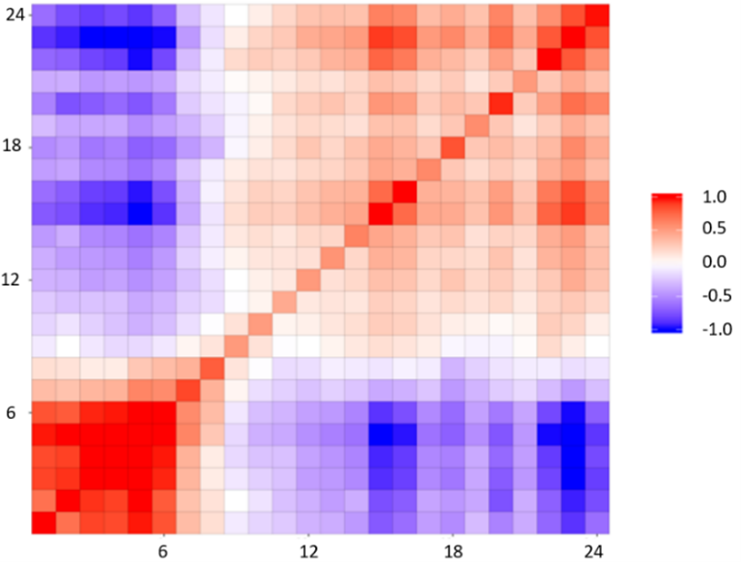}
	\caption{Covariance matrix of the multivariate normal random variable RT-LMP, for one sample day. }
	\label{fig:Cov_Matrix}
\end{figure}

In Fig. \ref{fig:Cov_Matrix} each pixel shows the covariance of forecast errors between two different forecast times and the diagonal of the matrix demonstrates the variance of each RT-LMP random variables. This visualization helps in estimating and capturing the dependence of forecast errors over the time. Following the steps of generating statistical scenarios proposed in \cite{Pinson2008}, we will generate a series of trajectory lines. which collectively represent a range of potential RT-LMP predictions over the forecast horizon, with associated probabilities. Fig. \ref{fig:Price_Forecast} shows the associated scenarios reflecting both the prediction uncertainty and the interdependence structure of predictions errors.

\section{LAC Simulation}\label{sec:LACsimulation}
To implement and test the LMP forecasts and the PSH LAC models described in the previous sections a high-performance unit commitment software, HIPPO, is used and further developed to perform the LAC simulations. HIPPO is co-developed by Pacific Northwest National Laboratory (PNNL), MISO and MIP solver vendor Gurobi to solve large-scale security constrained unit commitment (SCUC) and economic dispatch (SCED) problem for a day ahead (DA) market HIPPO \cite{High2016pan}. The LAC rolling window simulations is developed based on the existing HIPPO. The framework of the LAC simulation is included in this subsection.

\begin{figure}[!h]
	\centering
	\includegraphics[width=0.42\textwidth]{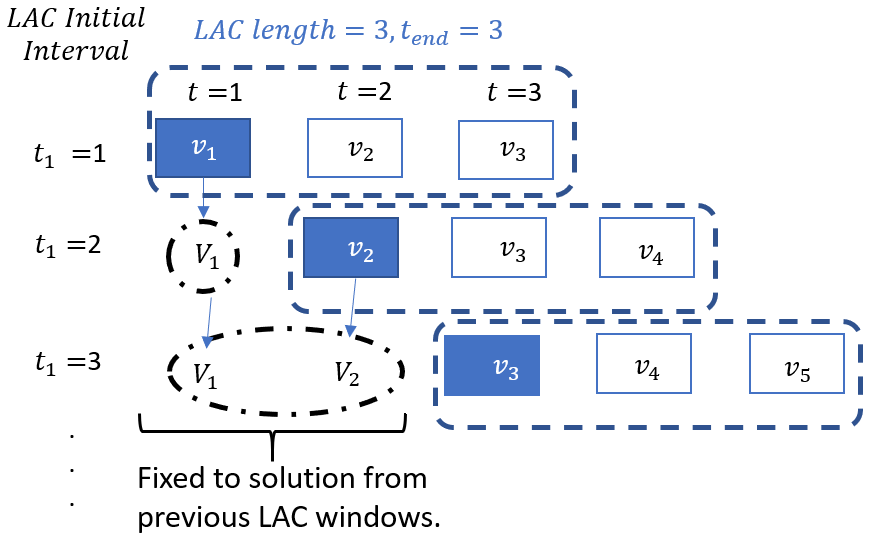}
	\caption{LAC Framework}
	\label{fig:lac_frame}
\end{figure}

%To keep minimum modifications to the model structures in HIPPO, each of the LAC rolling window is defined as follows: fix each of the variables outside the LAC window to a previously determined value and allow the variables inside the LAC window to be optimized. In this way, the SCUC or SCED problem is solved with every interval of the entire horizon $T$ for each rolling LAC window so that the constraints remain mostly unchanged. Although variables in every interval are included in the problem for each LAC window, the LAC problem can be solved fast since the only ``free'' variables are the ones inside the LAC window. Only the unit commitment variables after the LAC window are fixed to the DA solution such that the long lead units can satisfy all binary constraints. At the same time, some of the time-coupled constraints that link the variables after the LAC to the variables inside LAC such as ramping constraints are disabled to prevent major impacts from the fixed variables after the LAC. The system variables are indicated in the boxes in Fig \ref{fig:lac_frame}.

The framework of the LAC rolling window simulation in HIPPO is illustrated in Fig \ref{fig:lac_frame}. The system variables include the unit commitment and dispatch variables for generators and PSHUs $\mathcal{Q}$, $\mathcal{U}$, which are defined for each interval for the entire horizon in study: $v_t = \{\mathcal{Q},\mathcal{U} \}$,$\forall t \in T$. The LAC windows are highlighted by the dashed blue lines in Fig \ref{fig:lac_frame}. As an example, there are three intervals included in each LAC window in the figure but the number of intervals in a LAC window is a parameter and can be changed to any integer value between $1$ and the total number of intervals $T$. 
%Notice that the problem is solved with every interval of the entire horizon $T$ represented. However, the intervals after the LAC window highlighted by the dashed orange lines are fixed to a DA solution that is available before the LAC rolling simulation starts. 

Although LAC has sub-hour intervals in practice, we perform the LAC simulation with hourly intervals as a simplification. The hourly intervals allow straightforward comparison of results with DA solutions and it is easier to validate. An illustration of the LAC simulation is provided in Fig \ref{fig:lac_frame}. The first LAC problem starts at $t_1=1$ and it is indicated by the first row of the boxes representing variables in each of the intervals in Fig \ref{fig:lac_frame}. After the first LAC problem is solved, the solutions to the variables of the first interval inside the LAC window, that is $v_1$ written in white font and highlighted in the box filled with blue background, is saved and set as the fixed value $V_1$  to the variables in interval $1$ in the next and following LAC problems shown in the dot dashed black circle. The second LAC window starts at $t_1=2$ with the fixed solution from the previous window $V_1$. The LAC window slides forward one interval to $t \in [2,4]$. 
%the solution in the first interval inside the previous LAC window is fixed to the variables in the same interval of the current LAC problem. The first interval after the end of the previous LAC window, that is $v_4$ for LAC window $t_1=2$, is included in the current LAC window for optimization. After the second LAC window is solved, similarly, the solutions to the variables of the first interval inside the LAC window, that is $v_2$ written in white font and highlighted in the box filled with the blue background, is saved as $V_2$ . Along with the solution of the first interval from the first LAC window, $V_1$ and $V_2$ are set as the fixed values for the variables in intervals $1$ and $2$ in the next and following LAC problems shown in the dot dashed black ellipse. 
The LAC simulation rolls forward one interval at a time in the same way until the last interval inside the LAC window reaches the last interval of the entire horizon $T$.

\section{Numerical Studies}

In this section, first the system and PSHU data used in the LAC simulation is introduced. Then, four MISO production days are studied with five models including the stochastic PSH LAC model described in section \ref{sec:StoModel}, the robust model described in section \ref{sec:RobModel} and three benchmark models are presented. %\textcolor{blue}{The current PSHU practice model in the MISO real-time market is developed as the first benchmark. In addition, a Perfect PSH LAC model with full day look-ahead horizon is developed to gauge the performance of the proposed models. In the end, a deterministic model is presented as the third benchmark.}

%\textcolor{blue}{Because it is non-trivial to process system production data for large amount of days, a simple profit maximization simulation applying the price forecast are used to screen candidate days for study. The details of the profit maximization simulation can be found in \cite{ghesmati2022probabilistic}. Based on the simulation results, the four selected days that represent the RT profit potentials ranging from low to high are selected.}

While it is difficult to exactly reproduce the real-time (RT) system condition, we assembled the LAC simulation data from the production RT data and made necessary adjustment. The after-the-fact RT system wide demand, generator, security transmission constraints for each time interval is updated in each of the rolling LAC windows. It is noted that approximations are necessary to resolve some inconsistency and to attain feasible solutions. 
%In addition, it is found difficult to exactly reproduce the RT LMP trends from the simulation only with the RT system data because the existence of out-of-market actions and procedures in the RT operation. Therefore, the RT demand used in the simulation is slightly modified to recreate the after-the-fact RT LMP trends. 
Two PSH plants are included in this study. The parameters of the units are matched with production data. The capacities of the PSHU ranges from a few hundred MW to GW. Because the confidentiality of the production data, the detailed parameters of the PSH units cannot be presented in this paper.

\subsection{Current PSH Practice Model}
The first benchmark is current PSH practice in the LAC. In current operation, MISO does not optimize state of charge for PSH in LAC. The PSHU usually stay with their DA position to avoid risks in the RT market. Therefore, in this study, the proposed models are first compared to the current practice model.

\subsection{Perfect PSH LAC Model}

The second benchmark is the perfect LAC model where the PSHU is fully optimized in the day. That means all the unit constraints are fully represented in the system wide optimization including the unit output limits, ramp limits and SOC limits etc. The end of the day SOC is set to meet the end of the day target in the DA solution. The after the fact RT system conditions are known in the perfect model. Therefore, the perfect LAC model should provide the best performance benchmark for the other models.

\subsection{Deterministic PSH LAC Model}
The third benchmark is a deterministic PSH model. The deterministic PSH model is exactly the same as the stochastic model described in section \ref{sec:StoModel} except there is only one scenario of price forecast instead of many. The single point LMP forecast used in the deterministic PSH model is given by the methodology described in section \ref{sec:ARIMAX}. This benchmark is developed to study the impacts of including probabilistic price forecast in the PSH LAC simulation. 

\subsection{Simulation Results}

%In this study, the MISO system data in total four days are selected. Each includes over $1000$ generators. Reserve requirements and transmission security constraints are included for all studies. Constraints on individual generators such as, minimum up/down time, maximum start up time, ramp constraints are included for all units. For each studied day, a DA problem is solved first to provide the reference data for the risk management constraint \eqref{eq:riskmanage} and it is used to calculate profit as shown in \eqref{eq:LACprofit}. Then, the LAC simulation is solved five times with the five different models for the PSHU in each studied day. The five models include the three benchmark models introduced in this section and the stochastic and robust PSH model described in section \ref{sec:StoModel} and section \ref{sec:RobModel}. For a brief presentation, they are referred as The Current Practice, Perfect LAC, Deterministic, Stochastic and Robust in this section. All optimization problems are solved with Gurobi 8.0.

\begin{table}
	\caption{System Objective Compared to The Current Practice}
	\label{table:System Objective}
	\centering
	\begin{tabular}{c c c c c c}
		\hline
		Model & Day 1 & Day 2 & Day 3 & Day 4\\
		\hline
		Perfect LAC  & $-0.57\% $ & $-0.13\%$ & $-0.08\%$  & $-0.38\%$ \\
		Deterministic  & $-0.47\%$ & $+0.16\%$ & $-0.03\%$ & $-0.007\%$ \\
		Stochastic  & $-0.55\%$ & $+0.1\%$ & $-0.08\%$ & $-0.3\%$ \\
		Robust  & $-0.33\%$ & $+0.09\%$ & $-0.08\%$ & $-0.29\%$ \\
		\hline
	\end{tabular}
\end{table}

\begin{table*}[!h]
	\centering
	\caption{PSHU RT Profit as Percentage of DA Profit}
	\label{table:PSHU Profit}
	%\footnotesize
	\begin{tabular}{c c c c c c c c c}
		\hline
		\begin{tabular}[c]{@{}c@{}} \multirow{2}{*}{} Model \end{tabular}  & \multicolumn{2}{c}{Day 1} & \multicolumn{2}{c}{Day 2}  & \multicolumn{2}{c}{Day 3}  &\multicolumn{2}{c}{Day 4} \\
		& PSHU1 & PSHU2 & PSHU1 & PSHU2 & PSHU1 & PSHU2 & PSHU1 & PSHU2 \\ \hline
		Perfect LAC & $+16\%$ & $+9\%$ & $+2.0\%$ & $+0.9\%$ & $+14.76\%$ & $+2.9\%$& $+19.3\%$ & $+9.5\%$ \\ \hline
		Deterministic  & $+11\%$ & $+7\%$ & $-0.92\%$ & $-2.5\%$ & $+4.0\%$ & $+1.17\%$ & $+1.1\%$ & $+7.3\%$ \\ \hline
		Stochastic & $+13\%$ & $+9\%$ & $-2.1\%$ & $-1.1\%$ & $+13.6\%$ & $+2.0\%$ & $+2.3\%$ & $+5.6\%$ \\ \hline
		Robust & $+8\%$ & $+4\%$ & $+0.19\%$ & $-1.6\%$ & $+10.88\%$ & $+4.15\%$ & $+6.8\%$ & $+3.0\%$ \\ \hline
	\end{tabular}
\end{table*}

Shown in Table \ref{table:System Objective}, the system objective value of the studied models are compared. To show the comparison, the objective value of each of the rest of the models is compared to the current practice in Table \ref{table:System Objective}. Each cell gives the differences in percentage to the objective value of the current practice. A negative sign indicates a reduction of objective value compared to the current practice. Because the PSHU stays with the DA solution in LAC in the current practice instead of adapting to the RT system condition, it results in sub-optimality and high objective values. %\textcolor{blue}{Note that while there are penalty terms in the objective, there is no penalty term applied particularly to the PSHU constraints. Therefore, the reduction on the objective value compared to the current practice indicates system wide benefits that are composed by reduced generation cost and improved reliability.}
The Perfect LAC model uses the after the fact RT system information, therefore, as expected, it gives the lowest objective among all models for each of the studied days. Compared to the Deterministic model, the Stochastic model gives a lower system objective for every studied day. While the Robust model gives a higher objective than the Stochastic model in the Day $1$, the objective values of the two models are close in the rest of the studied days. It is observed that, except for Day $2$, the system objective is improved for all models. Due to the confidentiality of the system data, the actual system objective is not displayed. The range of the system objective value reduction is from multi-thousands dollars to multi-tens of thousands dollars. 

The PSHU profit realized in each model is presented in Table \ref{table:PSHU Profit}. After the LAC, if the PSHU deviates from its DA position, the PSHU would gain (or lose) profits from the RT market. The LAC profit is calculated as shown in \eqref{eq:LACprofit}. 

\begin{equation} 
	\begin{split}
		Profits_g^{LAC} = & LMP_{g,t}^{RT}[(Q_{g,t}^{gen,LAC}-Q_{g,t}^{pump,LAC})\\&-(Q_{g,t}^{gen,DA}-Q_{g,t}^{pump,DA})], \ \ \forall g \in \mathcal{G}_{psh}, 
	\end{split}
	\label{eq:LACprofit}
\end{equation}
where $LMP_{g,t}^{RT}$ is the post simulation RT LMP, $Q_{g,t}^{gen,LAC}$ and $Q_{g,t}^{pump,LAC}$ are the LAC dispatch solution for generation and pump modes. Because the PSHU stay with the DA solution in the Current Practice Model, it would not incur any non-zero profits in LAC. Therefore, the profit results of the Current Practice Model are not presented. 

In Table \ref{table:PSHU Profit}, the LAC profit of each of the models in a day is shown as a percentage to the DA profit of the unit in the same day. It can be observed that the Perfect LAC model always results in positive and the largest profits across all models in each of the studied days. The profit results from the Perfect LAC serve as indications of how much room left for the PSHU to adjust their positions and provide improvement in RT market. In Day $2$, the profit from the Perfect LAC is significantly less than the other three days. That gives a tight upper bound of how good each of the models can do in that day. In Day $3$ and Day $4$, compared to the Deterministic model, the Stochastic and Robust models generate significantly more profits for both PSHUs. In Day $2$, while the Stochastic models causes profit loss to both PSHUs, the Robust model could still generates a small amount of positive profit for PSHU1 and causes a relatively small profit loss to PSHU2.

% distribution of system cost and unit profits
%From the results shown in Table \ref{table:System Objective} and Table \ref{table:PSHU Profit}, it is observed that the distribution of system cost and unit profits are wider with the stochastic model. The robust model shrinks the distribution, preventing higher costs/profit lost but also decreasing the benefits in the other cases. Table I gives reference on the system objective reduction. In summary, the stochastic model will increase the system objective by $0.1\%$ in the worst case and decrease $0.55\%$ in the best case. In contrast, the robust model has a range of $-0.33\%$ to $+0.09\%$. The robust model shrinks the distribution of costs by $0.23\%$. It reduces higher costs by $0.01\%$ and it can also decrease the benefit of system objective reduction by up to $0.22\%$. Similar observation can be made from Table II which gives reference on the unit RT profits. If we take PSHU1 as an example, the profit (as percentage to the DA profit) with the risk-neutral (Stochastic) model ranges from $-2.1\%$ to $+13.6\%$. The risk-averse (Robust) model has the range of $+0.19\%$ to $+10.88\%$. It is clear that the stochastic model has the risks of losing profit in the RT market but achieve a higher profit in some cases. The robust model shrinks the distribution of profits by $4.53\%$. It prevents profit losing but also could decrease profits by up to $2.52\%$.

Based on the observation from both Table \ref{table:System Objective} and Table \ref{table:PSHU Profit}, we can conclude that in the days (Day $1$,$3$,$4$) where there is a larger potential improvement in RT market, except for the Perfect LAC, the Stochastic model perform consistently better than all the tested models. The Robust model provide less but close to the profits from the Stochastic model. In Day $2$ when there is less potential improvement in the RT market, the Robust model out performs the other models.

\begin{figure}[!h]
	\centering
	\includegraphics[width=0.42\textwidth]{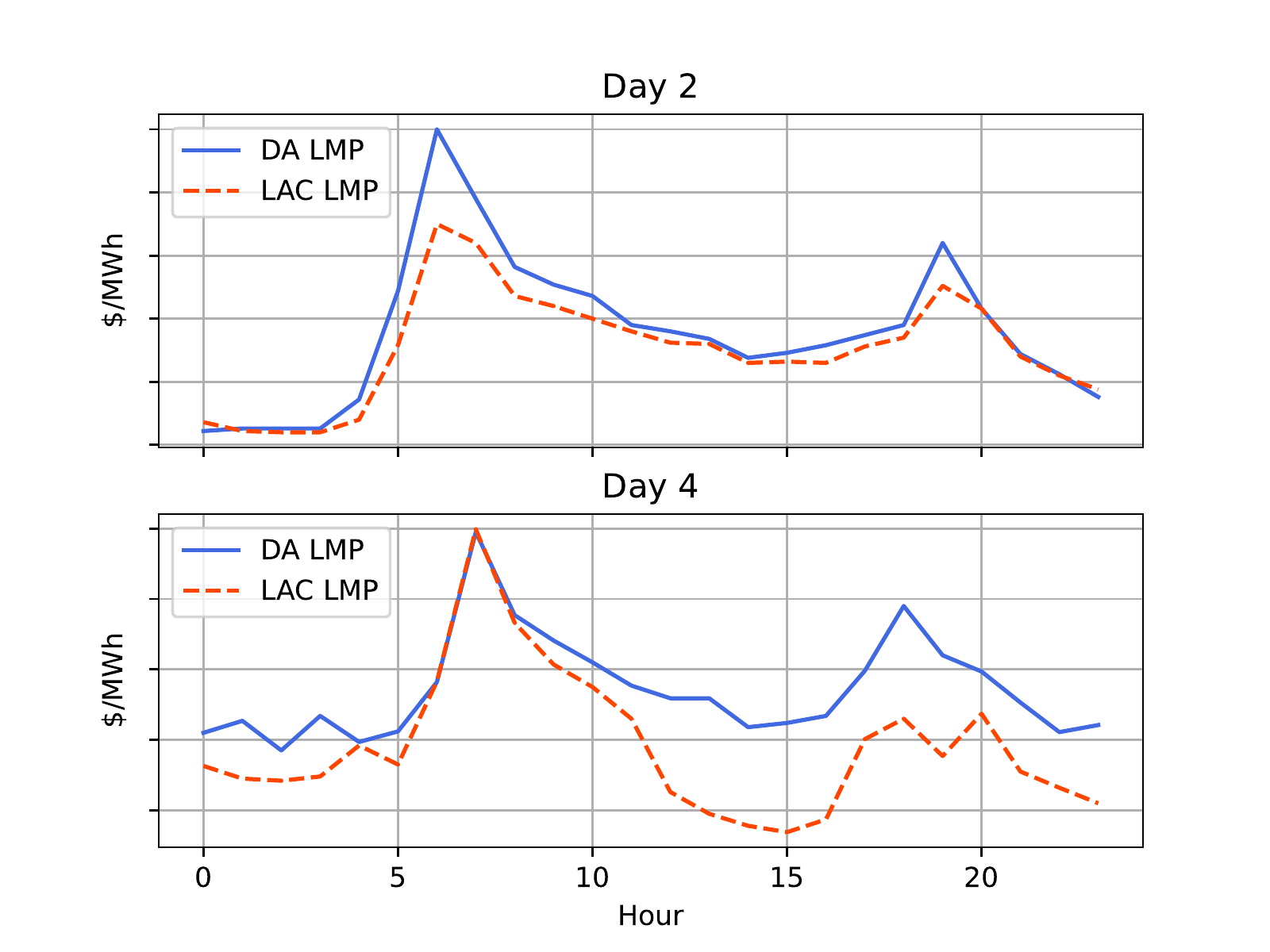}
	\caption{Simulated DA and LAC LMP in Day 2 and Day 4}
	\label{fig:DA and LAC LMP comparison Day 24}
\end{figure}

As an example, the post simulation DA LMP and LAC LMP in Day $2$ with less RT potential improvement and in Day $4$ with more RT potential improvement are illustrated in Fig. \ref{fig:DA and LAC LMP comparison Day 24}. In Day $2$, it can be observed that although the LAC LMP is lower than the DA LMP, the peak hours and the valley hours are the same. In comparison, in Day $2$, the evening peak shifted from hour $18$ in DA to hour $20$ in LAC. More importantly, the lowest LMP occurs at hour $15$ in LAC instead of hour $2$ in DA. Thus, in the days where the peak and/or valley hours are different in RT than they are in DA, it opens more opportunity to the PSHUs to adapt to that changes in LAC.

\begin{figure}
	\centering
	\includegraphics[width=0.5\textwidth]{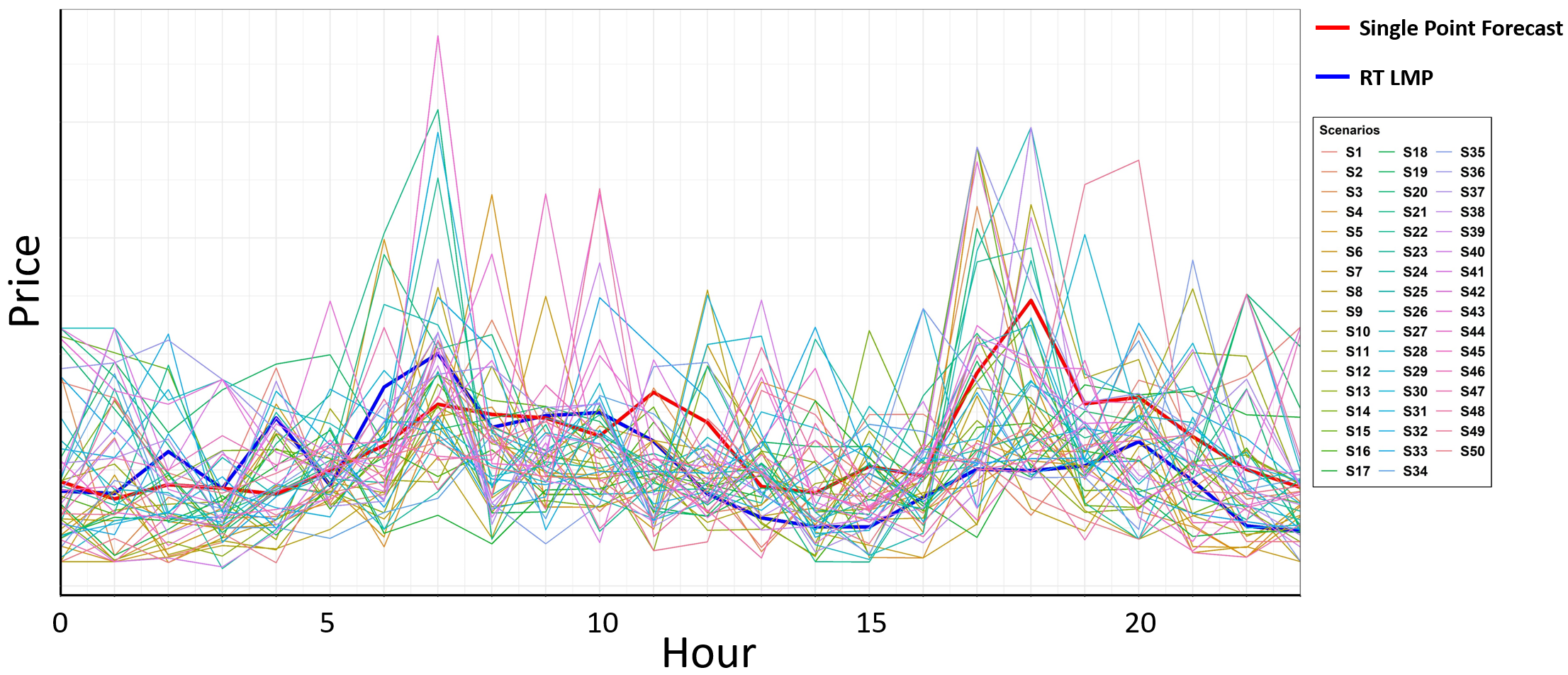}
	\caption{RT Price Forecast Results in Day 4}
	\label{fig:RT Price Forecast Results}
\end{figure}

The price forecast results made at hour $0$ are illustrated in Fig. \ref{fig:RT Price Forecast Results}. The single point forecast is shown as the bold red line, the after the fact RT LMP is shown as the bold blue line. The colored thin lines are the probabilistic price forecast scenarios. 
%Compared to the actual RT LMP in Fig. \ref{fig:RT Price Forecast Results}, although the simulated LAC LMP in Day 4 shown in Fig. \ref{fig:DA and LAC LMP comparison Day 24} is not exactly the same at each interval, the peak and valley hours and the trends in between are perfectly aligned.

\begin{figure}
	\centering
	\includegraphics[width=0.5\textwidth]{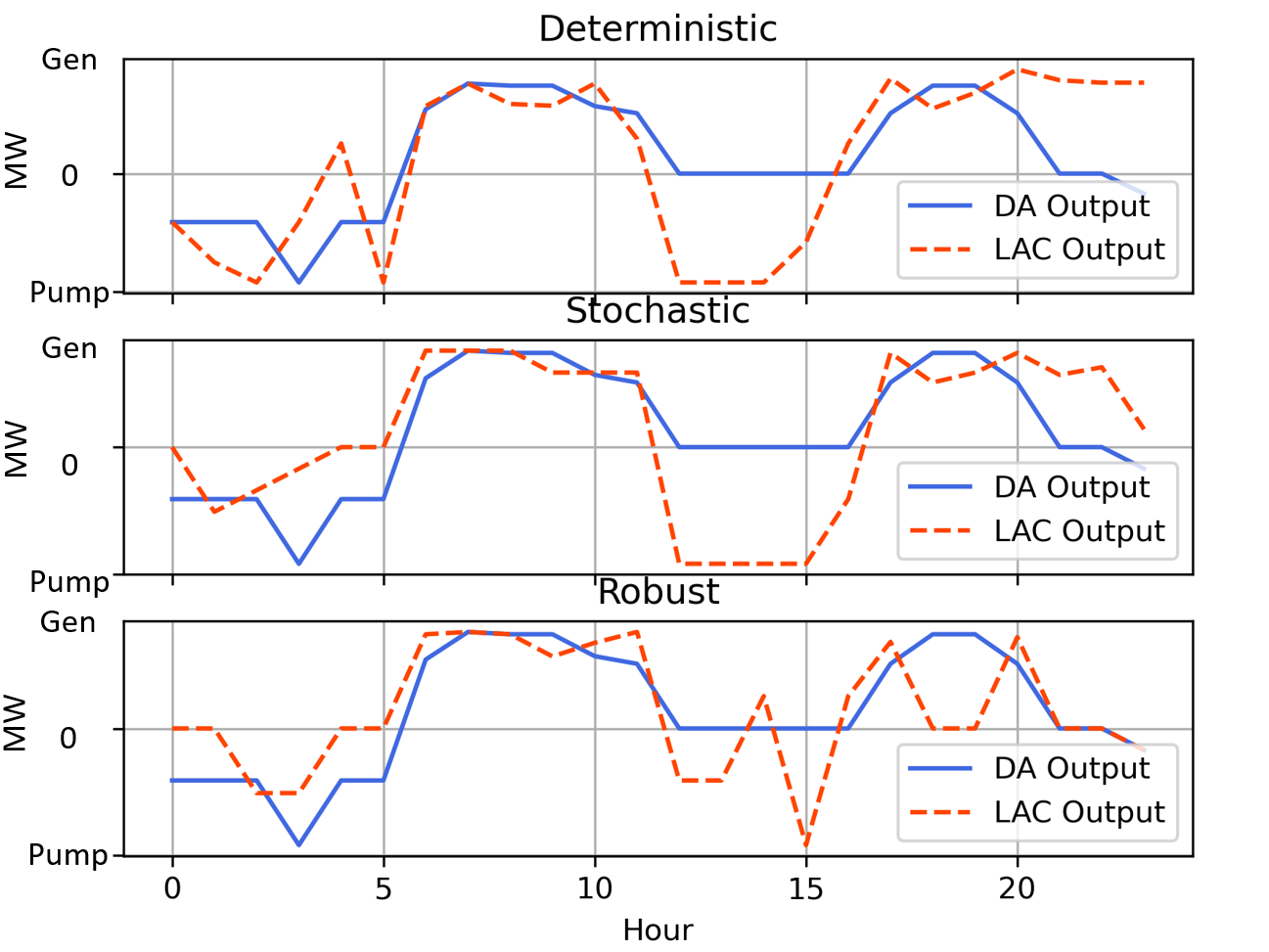}
	\caption{DA and LAC Dispatch Results for PSHU2 in Day 4}
	\label{fig:DA LAC Dispatch Results}
\end{figure}

The dispatch results of PSHU2 in Day 4 from Deterministic, Stochastic and Robust models are illustrated in Fig. \ref{fig:DA LAC Dispatch Results}. Shown in Fig. \ref{fig:RT Price Forecast Results}, the single point forecast doesn't predict the lower LMP in the afternoon, therefore the Deterministic model is not able to lead the PSHU to reduce the pump load in the early LAC windows. In contrast, in Fig. \ref{fig:RT Price Forecast Results}, it can be observed that there are a good coverage of probabilistic price scenarios in the actual valley time window around hour $15$. The effect of the probabilistic price forecast is demonstrated by the Stochastic and the Robust model reducing the pump load before hour $5$ from its DA position significantly. As a result, the Stochastic model successfully adjusts the PSHU's position in the series of LAC windows and shifts a good amount of pump load from the DA schedule in the early of the day to the afternoon around hour $15$ where the actual valley of the day appears. Due to the effect from the risk constraints, the output of the PSHU in Robust model is pushed closer to its DA position and it doesn't pump as much as the Stochastic model does in the actual valley hours. But the Robust model at least catches the lowest LMP of the day at hour $15$ to pump to its full capacity. 

%All three models capture the RT evening peak hour at hour $20$ and generate to its full capacity. This is benefited by the three hour LAC window but not the probabilistic price forecast because the PSHU generate to its full capacity in the Deterministic model even though the single point price forecast doesn't predict the evening peak correctly.     

\subsection{Computational Time and MIP Size}

\begin{table}[!h]
	\caption{Computational Time of the First LAC window}
	\label{table:Computational Time}
	\centering
	\begin{tabular}{c c c c c c}
		\hline
		Number of Scenarios & 10 & 20 & 50 & 75\\
		\hline
		Stochastic [sec]  & $4.4$ & $5.92$ & $11.88$ & $15.38$ \\
		Robust [sec] & $6.11$ & $7.52$ & $10.72$ & $16.86$ \\
		\hline
	\end{tabular}
\end{table}

In this section, we present the computational results with different number of price forecast scenarios on Day 3. With the proposed PSH models, the first LAC window includes the largest number of intervals that applies the price forecast. Thus, the problem size and the computational time of the first LAC window will be impacted the most by the price forecast scenarios. Therefore, we only compare the computational time and problem size for the first LAC window.

The first LAC in current practice that does not use any price forecast is solved in $3.13$ sec. Shown in Table \ref{table:Computational Time}, the robust model solves slower than the stochastic model and both model takes longer time to solve with more scenarios. Shown in \ref{table:Problem Size}, the problem size increase as the result of increased number of scenarios and that is consistent with the computational time results. However, even with the $75$ scenarios, both models can be solved within $20$ seconds and can meet the computational needs for real time LAC.

\begin{table}[!h]
	\caption{Increase Rate on the Problem Size in the First LAC window}
	\label{table:Problem Size}
	\centering
	\begin{tabular}{c c c c c c}
		\hline
		Number of Scenarios & 10 & 20 & 50 & 75\\
		\hline
%		Stochastic  &  &  & &  \\
		Rows  & $4.2\%$ & $9.1\%$ & $23.9\%$ & $36.2\%$ \\
		Columns  & $3.2\%$ & $6.8\%$ & $17.8\%$ & $26.8\%$ \\
		Non-zeros  & $5.7\%$ & $12.5\%$ & $32.8\%$ & $49.7\%$ \\
%		\hline
%		Robust  &  &  & &  \\
%		Rows  & $4.2\%$ & $9.1\%$ & $23.9\%$ & $36.2\%$ \\
%		Columns  & $3.2\%$ & $6.8\%$ & $17.8\%$ & $26.8\%$ \\
%		Non-zeros  & $6\%$ & $12.9\%$ & $33.9\%$ & $51.4\%$ \\
		\hline
	\end{tabular}
\end{table}

\section{Conclusion}
Due to the risk concerns, PSH owners currently avoid active participation in the RT market. To explore the potential opportunity to enhance market efficiency by including PSH in the RT market optimization, we propose a novel and practical Stochastic and a Robust model that uses probabilistic price forecast to adjust PSHU's positions in the RT rolling LAC. The proposed model only requires probabilistic price scenarios for PSHU in the intervals post to a LAC. This modeling design provides two major benefits. First, by avoiding taking stochastic scenarios input from load and renewable energy resources, it is computationally very efficient and does not impose challenges in the real-time operation. Second, only limited system parameters are required to change from the current production LAC model  and that greatly simplify the potential testing and implementation.

The findings from the MISO case studies can be summarized in three aspects. First, using probabilistic price forecast in the RT rolling LAC windows can help to account for uncertainties in future intervals outside of LAC window when adjusting PSHU's charging/discharging positions. Those adjustments can make significant improvement in the days when the RT system condition deviates from the forecast in DA. Second, the proposed robust model shows the effect of risk reduction particularly in the days when the RT system condition aligns better with the forecast in DA. Third, it is computationally efficient to use price forecast scenarios for the PSHUs in the LAC model. Both the stochastic and robust models performs well and can solve within $20$ seconds.

\ifCLASSOPTIONcaptionsoff
  \newpage
\fi



%\bibliographystyle{IEEEtran}
%\begin{thebibliography}{1}
%\end{thebibliography}

\bibliographystyle{IEEEtran}
\bibliography{LAC_PSH}


\iffalse
\begin{IEEEbiography}{Michael Shell}
Biography text here.
\end{IEEEbiography}

\begin{IEEEbiographynophoto}{John Doe}
Biography text here.
\end{IEEEbiographynophoto}

%\newpage

\begin{IEEEbiographynophoto}{Jane Doe}
Biography text here.
\end{IEEEbiographynophoto}
\fi

%\vfill


\end{document}